\documentstyle[12pt]{article}
\input{epsf.tex}
\def\lsim{\mathrel{\rlap {\raise.5ex\hbox{$ < $}}
{\lower.5ex\hbox{$\sim$}}}}
\def\gsim{\mathrel{\rlap {\raise.5ex\hbox{$ > $}}
{\lower.5ex\hbox{$\sim$}}}}
\def\thebibliography#1{%
\vskip 0.5cm \centerline{\bf References}
\list{%
[\arabic{enumi}]}{\settowidth\labelwidth{[#1]}
\leftmargin\labelwidth
\advance\leftmargin\labelsep
\usecounter{enumi}}
\def\newblock{\hskip .11em plus .33em minus .07em}
\sloppy\clubpenalty4000\widowpenalty4000
\sfcode`\.=1000\relax}

\def\Im{{\rm Im}\, }
\def\Re{{\rm Re}\, }
\def\np#1#2#3{Nucl. Phys. {\bf{B#1}} (#2) #3}
\def\pl#1#2#3{Phys. Lett. {\bf{#1B}} (#2) #3}
\def\prl#1#2#3{Phys. Rev. Lett. {\bf{#1}} (#2) #3}
\def\pr#1#2#3{Phys. Rev. {\bf{D#1}} (#2) #3}
\def\nl{\hfil\break}
\begin{document}
\begin{titlepage}
\begin{flushright}
CERN-TH/95-284\ \\
SISSA/170/95/EP\\
hep-th/9601037{\hskip.5cm}\\
\end{flushright}
\begin{centering}
\vspace{.3in}
{\bf UNIVERSAL MODULI-DEPENDENT STRING THRESHOLDS} \\
{\bf IN $Z_2\times Z_2$ ORBIFOLDS} \\
\vspace{3 cm}
{P.M. PETROPOULOS$^{\ 1,\, \ast}$
  and J. RIZOS$^{\ 1,\, 2,\, 3,\, \diamond}$}\\
\vskip 1cm
{$^1 $\it Theory Division, CERN}\\
{\it 1211 Geneva 23, Switzerland}\\
\bigskip
{$^2 $\it Division of Theoretical Physics, Physics Department,
University of Ioannina}\\
{\it 45110 Ioannina, Greece}\\
\medskip
{\it and}\\
\medskip
{$^3 $\it International School for Advanced Studies, SISSA}\\
{\it Via Beirut 2-4, 34013 Trieste, Italy}\\
\vspace{1.5cm}
{\bf Abstract}\\
\end{centering}
\vspace{.1in}
In the context of a recently proposed method for computing
exactly
string loop corrections regularized in the
infra-red, we determine and calculate the universal
moduli-dependent
part of the threshold corrections to the gauge couplings for the
symmetric
$Z_2\times Z_2$ orbifold model.  We show that these corrections
decrease the unification scale of the underlying effective field
theory. We also comment on the relation between this infra-red
regularization scheme and other proposed methods.
\vspace{1cm}

\noindent Published in Phys. Lett. {\bf B374}(1996)49.

\begin{flushleft} CERN-TH/95-284\\
SISSA/170/95/EP\\
December 1995  \\
\end{flushleft}
\hrule width 6.7cm \vskip.1mm{\small \small \small $^\ast$\ On leave
from {\it Centre National de
la Recherche Scientifique,} France.\\
 $^\diamond$\ Work supported by the EEC contract ref.
ERBCHBGCT940634.} \end{titlepage}
\newpage
String theories provide a unified description of gauge and
gravitational interactions at a scale close to the Planck mass
$M_P=\frac{1}{\sqrt{32\pi G_N}}$.
For four-dimensional vacua of the heterotic string
the relation between the running gauge coupling $g_\alpha(\mu)$
of the  low-energy effective field theory
and the string coupling $g_s$,
assuming the decoupling of massive modes,
must have the following form:
\begin{equation}
\frac{16\pi^2}{g_\alpha^2(\mu)} = k_\alpha\frac{16\pi^2}{g_s^2}+
b_\alpha
\log\frac{M_s^2}{\mu^2} + \Delta_\alpha \ ,
                                                       \label{one}
\end{equation}
where $b_\alpha$ are the usual effective field theory
beta-function coefficients of the
group factor $G_\alpha$, and $k_\alpha$ is the level of the
associated
Ka\v c--Moody algebra.
The thresholds $\Delta_\alpha$ are due to the
infinite tower of string modes and can be calculated at the level of
string theory \cite{VK, DKL, ANT, MOO, IGN, KK, KAH}. For
symmetric-orbifold models they have the general form:
\begin{equation}
\Delta_\alpha = -\hat{b}_\alpha \log\left(|\eta(T)|^4|\eta(U)|^4
\Im T \, \Im U\right) - k_\alpha Y(T,U)-
c_\alpha \ ,
\label{two}
\end{equation}
where $T$, $U$ are the vev's of gauge singlet fields corresponding
to the moduli of the internal torus \footnote{\ Here
$T_i = T$, $U_i = U$. } and $\hat{b}_\alpha$ are the
($N = 2$)-sector contributions to the beta-function coefficients.
The term $Y(T,U)$ stands for the universal
group-independent contribution to the threshold corrections,
and $c_\alpha$ are group-dependent constants. These constants are
also
scheme dependent, that is, they depend on the renormalization scheme
in
which
the running gauge couplings ${g_\alpha(\mu)}$ are defined.

String unification relates the fundamental string scale
$ M_s\equiv\frac{1}{\sqrt{\alpha'}} $ to the Planck scale and to the
string
coupling constant $g_s$ which is associated with the
expectation value
of the dilaton field. At the tree level this relation reads
\begin{equation}
M_s = g_s M_P\label{msmp}\ .
\label{three}
\end{equation}
Given the fact that low-energy data, assuming the minimal
supersymmetric  standard model as
the underlying low-energy field   theory, indicate
gauge  unification at a scale $M_X\sim 2\times 10^{16}$ GeV
\cite{AET}
which is two orders of magnitude less than the Planck scale,
threshold
corrections (\ref{two}) play a crucial role in string unification.
Their effect has been extensively studied in the literature
\cite{ILR,LAM,KPR,NIL,MNS,DF}
\footnote{\ Of course, alternative  possibilities
are either to modify the low-energy spectrum in order to increase the
effective field theory unification scale \cite{DF,POL}, or to choose
a
non-standard hypercharge normalization \cite{NSH}.} except for the
moduli-dependent universal terms $Y(T,U)$
which have received little attention because they
can be formally reabsorbed into a redefinition of ${g_s}\,$.
However, such a redefinition alters eq. (\ref{three}) in a
moduli-dependent way and consequently the relation between
string unification scale and Planck mass gets modified.
Following this observation, the purpose of the present letter is to
evaluate explicitly these terms in the context of the symmetric
 $Z_2\times
Z_2$ orbifold model and study their effect on the unification
of the gauge
couplings. We choose this model for   several
(related) reasons: (\romannumeral1) there are no one-loop corrections
to the
relation (\ref{three}) between the Planck scale and the string
coupling \cite{KK, KKA},
(\romannumeral2) there is no Green-Schwarz-like threshold
$\Delta^{\scriptstyle\rm univ}(T,U)$ i.e. no axion-dilaton-moduli
mixing at
the one-loop level, and
(\romannumeral3) there are no ($N = 1$)-sector contributions to the
beta-functions ($\hat{b}_\alpha
= b_\alpha$); the last two points allow us to define
the unification scale in a
manifest way. As we will see in the sequel, the
contribution of the group-factor   independent terms $Y(T,U)$ has a
decreasing effect on the unification scale. Besides the relevance
that
the universal contributions $Y(T,U)$ might have in the string
unification,
we should mention that their non-harmonic part is also related to
the one-loop correction of the K\"ahler potential for the
 moduli fields \cite{KAH, COR}. The K\"ahler potential turns out
 to be related to the one-loop corrections of the Yukawa
 couplings, and its moduli dependence has attracted much attention
 in the rapidly growing subject of dualities \cite{DUA}.

In a recent article
\cite{KK} an interesting method for computing
unambiguously the string loop corrections
has been proposed. The procedure which is used
consists of replacing flat four-dimensional space-time with a
suitably
chosen curved one in a way that preserves gauge symmetries,
supersymmetry and modular invariance, and with curvature that
induces an infra-red cut-off and can be
consistently switched off.
In this framework, vertices for space-time fields such as
$F_{\mu\nu}^\alpha$
are truly marginal world-sheet operators and therefore deformations
induced
by the associated background fields are exactly calculable. This
allows
in particular for the computation of various one-loop correlators.
For instance, in the case of the symmetric orbifolds, the
vacuum amplitude with two insertions of the magnetic background
operator reads
\begin{equation}
Z_2^\alpha\left(\frac{\mu}{M_s}\right) =
\int_{\cal F}\frac{d^2\tau}
{\Im \tau}\Gamma\left(\frac{\mu}{M_s}\right)
\sum_{i=1}^3
\frac{\Gamma_{2,2}(T_i,U_i)}
{\overline{\eta}^{24}}\left[{\overline{Q}}_\alpha^2
 - \frac{k_\alpha}{4\pi \Im\tau}\right]\overline{\Omega} \ .
\label{four}
\end{equation}
Here ${\Gamma_{2,2}(T_i,U_i)}$ are the internal two-torus
solitonic contributions
and
${\overline{Q}_\alpha}$ is the charge operator
associated with the group factor $G_\alpha$, its square acting as
$\frac{i}{\pi}
\frac{\partial}{\partial\overline{\tau}}$  on the appropriate part of
the model-dependent function $\overline{\Omega}$. The latter is a
modular
function of degree $10$. It deserves stressing here that the
radiative corrections (\ref{four}) include exactly the
back-reaction of the gravitationally coupled fields;
this accounts for the term
\footnote{\ This universal term was also found in \cite{IGN} for the
three-point function of two gauge bosons and the modulus~$T$.
%This correlation function leads to the exact expression of
%$\partial \Delta_{\alpha}/\partial T$ but does not enable to
% determine
%unambiguously $\Delta_{\alpha}$.
}
$\frac{-k_\alpha}{4 \pi \Im\tau}$ which
is universal and guarantees modular invariance.
Finally, the (suitably normalized) $SU(2)_k$ partition function
\begin{equation}
\Gamma(x) = \left.-2 x^2\sqrt{\Im\tau}\frac{\partial}{\partial x}
\left[Z(x) - Z(2x)\right]\right|_{x=\frac{1}{\sqrt{k+2}}}
\label{dreg}
\end{equation}
with
\begin{equation}
Z(x) = \sum_{m,n}e^{\frac{i\pi\tau}{2}\left(mx+\frac{n}{x}\right)^2}
e^{-\frac{i\pi\overline{\tau}}{2}\left(mx-\frac{n}{x}\right)^2}
\label{zz}
\end{equation}
replaces the flat-space contribution and ensures the convergence of
the
integral
at large values of $\Im\tau$ by introducing a universal mass gap
$\mu=\frac{M_s}{\sqrt{k+2}}$
to all string excitations. This infra-red regularization of the
string
(on-shell)
loop amplitude vanishes when the flat-space limit is reached since
$\lim_{x\to 0}\Gamma(x) = 1\,$.

Before going further in the evaluation of the group-independent terms
$Y(T,U)$
we would like to relate the above correlator (\ref{four}) to the
one-loop
effective
field theory gauge couplings $g_\alpha(\mu)$ renormalized at some
scale
$\mu$,
in a specific ultra-violet scheme,  the ${\overline {DR}}$ scheme,
and
show that despite the presence of the curvature-induced infra-red
regulator
$\Gamma(\frac{\mu}{M_s})$ in (\ref{four}), the thresholds are
infra-red-cutoff independent. This can be done by identifying the
string
one-loop corrected  coupling
\begin{equation}
k_\alpha\frac{16\pi^2}{g_s^2} +
Z_2^\alpha\left(\frac{\mu}{M_s}\right) ,
\label{five}
\end{equation}
with the corresponding field theory one-loop corrected gauge
coupling,
similarly
regularized in the infra-red
\begin{equation}
\frac{16\pi^2}{g_{\alpha,{\scriptstyle\rm bare}}^2} +
b_\alpha(4\pi)^\epsilon\int_0^\infty\frac{dt}{t^{1-\epsilon}}
\Gamma_{FT}\left(\frac{\mu}{\sqrt{\pi} M}\right)\ .
\label{six}
\end{equation}
Here we have used dimensional regularization for the ultra-violet and
$M$ is an arbitrary mass
scale; $b_\alpha$ are the full beta-function coefficients for the
group
factor
$G_\alpha$ and $\Gamma_{FT}$ is the field theory counterpart of the
infra-red regulator
(\ref{dreg}), obtained by dropping all winding modes
\footnote{\ The extra $\sqrt{\pi}$ in the
argument of $\Gamma_{FT}$
accounts for the identification of the (dimensionless in the above
convention)
Schwinger proper-time parameter $t$ with $\pi\Im\tau\,$.}. On the
other hand, one knows
that in the $\overline{DR}$ scheme the relation between the field
theory bare
and running coupling is
\begin{equation}
\frac{16\pi^2}{g_{\alpha,{\scriptstyle\rm bare}}^2} =
\frac{16\pi^2}{g_{\alpha}^2(\mu)} -
b_\alpha(4\pi)^\epsilon\int_0^\infty\frac{dt}
{t^{1-\epsilon}}e^{-t\frac{\mu^2}{M^2}}.
\label{seven}
\end{equation}
Plugging (\ref{seven}) into (\ref{six}) and identifying the latter
with
(\ref{five}) leads to
\begin{equation}
\frac{16\pi^2}{g_{\alpha}^2(\mu)} =
k_\alpha\frac{16\pi^2}{g_s^2} +
Z_2^\alpha\left(\frac{\mu}{M_s}\right)
- b_\alpha(2\gamma+2)\ .
\label{KK}
\end{equation}

Equation (\ref{KK}) has been obtained by using an explicitly
infra-red-regularized string loop
amplitude. It is worthwhile to compare this expression to the one
derived in \cite{VK}
without any infra-red regulator. In order to do  so we first isolate
the
contribution of the massless
states
\begin{equation}
\lim_{\Im\tau\to\infty}\sum_{i=1}^3\frac{\Gamma_{2,2}(T_i,U_i)}
{\overline{\eta}^{24}}{\overline{Q}}_\alpha^2\,
\overline{\Omega}\equiv
b_\alpha^{\phantom 2}\ ,
\label{nine}
\end{equation}
responsible for the non-trivial infra-red behaviour of the integral
in
(\ref{four}). We then
rewrite (\ref{KK}) in the form
$$
\frac{16\pi^2}{g_{\alpha}^2(\mu)} =
k_\alpha \frac{16\pi^2}{g_s^2}
+
\int_{\cal F}\frac{d^2\tau} {\Im \tau}
\Gamma\left(\frac{\mu}{M_s}\right)
\left(\sum_{i=1}^3\frac{\Gamma_{2,2}(T_i,U_i)}
{\overline{\eta}^{24}}
\left[{\overline{Q}}_\alpha^2
- \frac{k_\alpha}{4\pi \Im\tau}\right]
\overline{\Omega}
- b_\alpha^{\phantom 2}\right)
$$
\begin{equation}
+ b_\alpha
\int_{\cal F}\frac{d^2\tau}
{\Im \tau}\Gamma\left(\frac{\mu}{M_s}\right)
- b_\alpha(2\gamma+2)
\label{eight}
\end{equation}
where we have subtracted and added back a $b_\alpha\int_{\cal
F}\frac{d^2\tau}
{\Im \tau}\Gamma(\frac{\mu}{M_s})$ term, a manipulation perfectly
well
defined
thanks to
the presence of the regulator $\Gamma(\frac{\mu}{M_s})$.
Inserting the result
\begin{equation}
\int_{\cal F}\frac{d^2\tau}
{\Im \tau}\Gamma\left(\frac{\mu}{M_s}\right) =
\log\frac{M_s^2}{\mu^2}
+ \log\frac{2e^{\gamma+3}}{\pi\sqrt{27}}+{\cal O}
\left(\frac{\mu}{M_s}\right)
\label{thirteen}
\end{equation}
into (\ref{eight}) and taking the limit $\mu\to0$ in the remaining
integral since it does not suffer any longer
from divergences at $\Im\tau\to\infty\,$, we finally obtain
$$
\frac{16\pi^2}{g_{\alpha}^2(\mu)} =
k_\alpha \frac{16\pi^2}{g_s^2} +
 \int_{\cal F}\frac{d^2\tau}
{\Im \tau}\left(\sum_{i=1}^3\frac{\Gamma_{2,2}(T_i,U_i)}
{\overline{\eta}^{24}}
\left[{\overline{Q}}_\alpha^2
- \frac{k_\alpha}{4\pi \Im\tau}\right] \overline{\Omega}
- b_\alpha^{\phantom 2}\right)
$$
\begin{equation}
+ b_\alpha\log\frac{M_s^2}{\mu^2}
+ b_\alpha \log\frac{2e^{1-\gamma}}{\pi\sqrt{27}} \ .
\label{eleven}
\end{equation}
As far as the group-factor dependent terms are concerned, expression
(\ref{eleven}) agrees, including the constant contribution, with the
one obtained in \cite{VK}  also in the $\overline{DR}$ scheme. Hence,
the relation between the running gauge couplings of the low-energy
field theory and the string coupling does not depend on the infra-red
regularization prescription. This result could have been anticipated
as a consequence of the cancellation of the infra-red divergences
between the fundamental and the effective theory since they have the
same massless spectrum. However it could only be proved in the
presence of a consistent infra-red regulator, similar in both
theories. Moreover, it is important to emphasize that (\ref{eleven})
contains rigorously all universal terms that were missing in
previous approaches
\cite{VK, DKL}
and that we will now determine.

The symmetric $Z_2\times Z_2$ orbifold model has a
$E_8\times E_6\times U(1)^2$
gauge group \footnote{\ In this case $k_\alpha =1$ and
$\hat{b}_\alpha
= b_\alpha$ with $b_{E_8}=-90$ and $b_{E_6}=126$.} and
$\overline{\Omega} =
\overline{\Omega}_8
\overline{\Omega}_6$
where
\begin{equation}
{\Omega }_8^{\phantom 8}=
\frac{1}{2}\sum_{a,b}{\vartheta}^8_{\phantom 8 }\left[^{a}_b\right]\
,\    \
{\Omega}_6^{\phantom 4}=\frac{1}{4}
\left({\vartheta}_2^4 + {\vartheta}_3^4\right)
\left({\vartheta}_3^4 + {\vartheta}_4^4\right)
\left({\vartheta}_2^4 - {\vartheta}_4^4\right) \ .
\end{equation}
These are proportional to the Eisenstein modular-covariant functions
${G}_2$ and ${G}_3$ respectively \cite{SLA}.
They are both related to the modular invariant ${j}$
(we use the standard normalizations, namely
$j(\tau)=\frac{1}{q}+744+{\cal O}(q)$, $q=\exp (2\pi i \tau)$):
\begin{equation}
\left(
\frac{\Omega_8}{\eta^8}
\right)^3 = j
\ , \ \
\left(
\frac{\Omega_6}{\eta^{12}}
\right)^2 = \frac{1}{4}\big(j - j(i)\big)
     \label{a}
\end{equation}
with $j(i)=12^3$.
The operator $\overline{Q}^2_{\alpha}$ acts as
$\frac{i}{8\pi} \frac{\partial}{\partial\overline{\tau}}$
on
$\overline{\Omega}_8$ for $\alpha =E_8$
while for $\alpha =E_6$ it acts similarly on the factors
${\overline{\vartheta}}^8_2$,
${\overline{\vartheta}}^8_3$ and ${\overline{\vartheta}}^8_4$
of $\overline{\Omega}_6^{\phantom 8}$.
One can use the above relations (\ref{a})
as well
as~\footnote{\ Eq. (\ref{b}) can be proved very easily. Both sides
are
holomorphic modular-invariant functions that have the
same analyticity properties. Therefore they must be proportional.
The proportionality factor is found by comparing the corresponding
power expansions.}
%(see \cite{SLA})
\begin{equation}
\frac{1}{\eta ^{4}}
\frac{\partial \log j}{\partial \log q} =
\frac{\left(j-j(i)\right)^\frac{1}{2}} {j^\frac{1}{3}}
\label{b}
\end{equation}
to show that
\begin{equation}
\frac{1}
{\overline{\eta}^{24}}\left[{\overline{Q}}_\alpha^2
- \frac{1}{4\pi \Im\tau}\right]\overline{\Omega}=
\frac{b_{\alpha}}{3} + 2
\left[{\overline{\Omega }}_2^{\phantom 2}
- \frac{1}{8\pi \Im\tau}\right]
\frac{\overline{\Omega}}{\overline{\eta}^{24}}
+ \frac{\overline{j}}{24}-42 \ ,
\label{c}
\end{equation}
where
\begin{equation}
\Omega _2=\frac{\partial \log \eta}{\partial \log q}
\label{d}
\end{equation}
is a non-modular-covariant function which plays a role in string
gravitational anomalies. Notice, however, that the non-holomorphic
combination $\Omega _2 - \frac{1}{8\pi \Im\tau}$
is modular covariant of degree $2$; once multiplied by
$\frac{\Omega}{\eta^{24}}$,
the latter is proportional to the gravitational $R^2$-term
renormalization.
If we introduce (\ref{c}) into (\ref{eleven}), we can perform the
integral corresponding to the group-factor dependent part by using
the result (for $T_i = T$, $U_i = U$)
\begin{equation}
\int_{\cal F}\frac{d^2\tau}
{\Im \tau}\left(\Gamma_{2,2}(T,U)
-1 \right) =
-\log\left(|\eta(T)|^4|\eta(U)|^4
\Im T \, \Im U\right) - \log\frac{8\pi e^{1-\gamma}}{\sqrt{27}} \ ,
\label{e}
\end{equation}
first established in \cite{DKL} and recently generalized in
\cite{MOO}.
Finally, comparison with eqs. (\ref{one}) and (\ref{two}) leads to
the universal part of the thresholds for the
$Z_2\times Z_2$ orbifold model,
\begin{equation}
Y(T,U) = \int_{\cal F}\frac{d^2\tau}
{\Im \tau} \Gamma_{2,2}(T,U) \left(-6
\left[{\overline{\Omega }}_2^{\phantom 2}
- \frac{1}{8\pi \Im\tau}\right]
\frac{\overline{\Omega}}{\overline{\eta}^{24}}
- \frac{\overline{j}}{8}+126
\right)\ ,
\label{twelve}
\end{equation}
as well as to the constants $c_\alpha$ in the $\overline{DR}$ scheme:
\begin{equation}
c_\alpha =  b_\alpha\log4\pi^2\ .
\label{con}
\end{equation}

A few remarks are in order here. We observe that, apart from
the expected universal threshold induced by the back-reaction term
$\frac{-1}{4\pi \Im\tau}$, there are other universal contributions
originated by the group-trace factor
$\frac{1}
{\overline{\eta}^{24}}{\overline{Q}}_\alpha^2
\overline{\Omega}$.
It is quite remarkable that the only group-factor dependence of the
latter
(see (\ref{c})) is a constant proportional to the beta-function
coefficients
while the other pieces are all universal. These play a very specific
role
and could almost have been guessed: combined with the back-reaction
term they
ensure modular invariance and finiteness of $Y(T,U)$ everywhere in
the moduli
space. Indeed, by using eqs. (\ref{a}) and the Fourier
expansion of $j$ it appears that
$6\frac{{\overline{\Omega
}}_2\overline{\Omega}}{\overline{\eta}^{24}}
=-\frac{1}{8}\frac{1}{\overline{q}}+33+{\cal O}(\overline{q}) $
and
$\frac{\overline{j}}{8}-126=
\frac{1}{8}\frac{1}{\overline{q}}-33+{\cal O}(\overline{q}) $.
The cancellation of constant and tachyonic terms avoids large-$\Im
\tau$
divergences in (\ref{twelve}) even when the gauge group gets
enlarged.

Expression (\ref{twelve}) can be further simplified if one uses a
generalization of (\ref{e}) (see \cite{MOO}), valid for more general
modular-invariant functions, to integrate the last terms
\footnote{\ Using the method of orbits of the modular group, the
remaining
integral in (\ref{f}) can also be reduced to a multiple series
expansion
\cite{MOO} but such a manipulation is not useful for our purpose.}:
\begin{equation}
Y(T,U) = \int_{\cal F}\frac{d^2\tau}
{\Im \tau} \Gamma_{2,2}(T,U)
\left(-6\left[{\overline{\Omega }}_2^{\phantom 2}
- \frac{1}{8\pi \Im\tau}\right]
\frac{\overline{\Omega}}{\overline{\eta}^{24}}
+33
\right)+\frac{1}{2}
\log|j(T)-j(U)| \ .
\label{f}
\end{equation}
Hence, while $Y(T,U)$ is finite and remains
finite along the whole line of enhanced symmetry $T=U$, both
terms of (\ref{f}) are logarithmic divergent when
$T\to U$ as a consequence of the extra massless states. This
divergence is precisely the one that is responsible for the
non-trivial monodromy properties of the prepotential in $N=2$
models \cite{MOO, FER, CAR}.

Finally, it is interesting to note that eqs. (\ref{a}) and (\ref{b})
allow us to recast the function $\Omega$ into the form:
\begin{equation}
\frac{\Omega}{\eta ^{24}}=\frac{j-j(i)}{2}
\left(\frac{\partial \log j}{\partial \log q}\right)^{-1}\ .
\label{ign}
\end{equation}
Although this expression strictly holds for the model under
consideration,
it seems that its validity could be extended (up to a factor) to more
general
string vacua with $N=2$ supersymmetry \cite{FER}. Again, advocating
modular
invariance and infra-red finiteness, one could draw the conclusion
that for
these models the universal thresholds are proportional to those of
the
$Z_2\times Z_2$ orbifold. At the present stage of investigation,
however,
this observation is by no means to be considered as a claim.

Let us now proceed to the numerical evaluation of
$Y(T,U)$ for the model at hand. We will concentrate
on the case $\Re T = \Re U = 0$
\footnote{\ We have verified numerically that as far as the
universal thresholds are concerned, switching on
the  $\Re T$ and $\Re  U$ fields leads to small dumping oscillations
(for increasing radii) with maximum width $<3\%$ around the  $\Re T =
\Re   U = 0$ results.}
and express $\Im T$, $ \Im U$ in terms of the internal radii:
$\Im T=R_1 R_2\, $, $\Im U=R_2/R_1$. Our starting point is
eq.  (\ref{twelve}) which reads now
\begin{equation}
Y(R_1,R_2) = \int_{\cal F}\frac{d^2\tau}{\Im \tau}
\Gamma_{1,1}(R_1)\, \Gamma_{1,1} (R_2) \,
\overline{\rho}
\label{yyr}
\end{equation}
where
$\Gamma_{1,1}(R)$ is the soliton contribution of a compactified
single
boson,
\begin{equation}
\Gamma_{1,1}(R) = \sum_{m,n} e^{-\pi\tau_2\left(\frac{m^2}{R^2}+n^2
R^2\right)}
e^{2\pi i\tau_1 m n}
\label{sol}
\end{equation}
(we set $\tau_1\equiv \Re \tau$, $\tau_2\equiv \Im \tau$), and
\begin{equation}
\overline{\rho} =
-6\left[{\overline{\Omega }}_2^{\phantom 2}
- \frac{1}{8\pi \Im\tau}\right]
\frac{\overline{\Omega}}{\overline{\eta}^{24}}
- \frac{\overline{j}}{8}+126 \ .
\end{equation}

One can readily derive the large-radius behaviour of the universal
corrections. For $R_1 = R_2 = R \gg 1$ we can neglect the windings,
setting $n=0$ in eq. (\ref{sol}), and split the integral (\ref{yyr})
into two terms:
\begin{equation}
Y(R,R) =
\int_{{\cal F}_1}\frac{d\tau_1 \, d\tau_2}{\tau_2}
\vartheta_3^2\left(i\frac{\tau_2}{R^2}\right)\, \overline{\rho}
+
\int_1^{\infty}\frac{d\tau_2}{\tau_2}
\vartheta_3^2\left(i\frac{\tau_2}{R^2}\right)\,
\int_{-\frac{1}{2}}^{+\frac{1}{2}}d\tau_1 \,\overline{\rho}
+ {\cal O} \left( e^{-\pi R^2} \right)  \ ,
                                       \label{yspl}
\end{equation}
where ${\cal F}_1$ corresponds to the part of the fundamental domain
with $\tau_2 < 1$. We can now perform a Poisson resummation in the
first term. Then, by using the results
\begin{equation}
\int_{{\cal F}_1}\frac{d\tau_1 \, d\tau_2}{\tau_2^2}
\overline{\rho}
= 6\pi-\frac{45}{\pi}
\end{equation}
and
\begin{equation}
\int_{-\frac{1}{2}}^{+\frac{1}{2}}d\tau_1\,\overline{\rho}
= \frac{90}{\pi \tau_2} \ ,
\ \
\int_1^{\infty}\frac{dx}{x^2}
\vartheta_3^2\left(i\frac{x}{R^2}\right) =
\frac{R^2}{2} + \frac{\kappa}{R^2} +
{\cal O} \left( e^{-\pi R^2} \right)
\end{equation}
in the first and second terms of (\ref{yspl}) respectively,
we finally obtain:
\begin{equation}
Y(R,R) =   6\pi R^2 +
\frac{90 \kappa}{\pi R^2} +
{\cal O} \left( e^{-\pi R^2} \right)
\end{equation}
where
$\kappa
=\frac{2}{\pi ^2} \zeta(4) + \sum_{j>0}
\left(
\frac{\coth \pi j}{\pi} \frac{1}{j^3} +
\frac{1}{\sinh^2 \pi j} \frac{1}{j^2}
\right)
$ is a constant that takes the numerical value
$\kappa \approx 0.6106\,$.
Thus the asymptotic form reads:
\begin{equation}
Y(R,R) \sim 18.85 \times R^2 +
17.49 \times \frac{1}{R^2}
\ \ {\rm for} \ \   R \gg 1 \ .
\label{yas}
\end{equation}

It is possible to compute explicitly the universal thresholds for
arbitrary
values of the radii $R_1, R_2\,$  by numerically evaluating
(\ref{yyr}).
Problems related to the loss of numerical accuracy caused by the pole
part $e^{2\pi\tau_2}$ in $\overline{\rho}(\tau_1,\tau_2)$ can be
cured by
power expanding the integrand  to the required accuracy, before
performing the numerical integration \cite{MNS}. Plots of the
numerical results
for $Y(R_1,R_2)$ as a function of $R_2$ for $R_1 = 1,2,3,4$ are given
in
fig.~1 and a contour plot of $Y(R_1,R_2)$ is given in fig.~2. One
clearly
sees that the minimum value of $Y$ is obtained for the radii at the
self-dual
point and it is $Y^{\scriptstyle\rm min} = Y(1,1) \approx 36.4$ while
for the
fermionic point we have $Y^{\scriptstyle\rm fermionic} =
Y(1,\frac{1}{2})
\approx 53.0\,$.
We also notice that the asymptotic formula
(\ref{yas})
reproduces the numerical results for $R_1 = R_2 \gsim \frac{3}{2}\,$,
which is consistent with the fact that
(\ref{yas})
is almost invariant under the duality transformation $R\rightarrow
1/R$.
\begin{centering}
\begin{figure}
\leavevmode
\epsfbox[80 240 570 540]{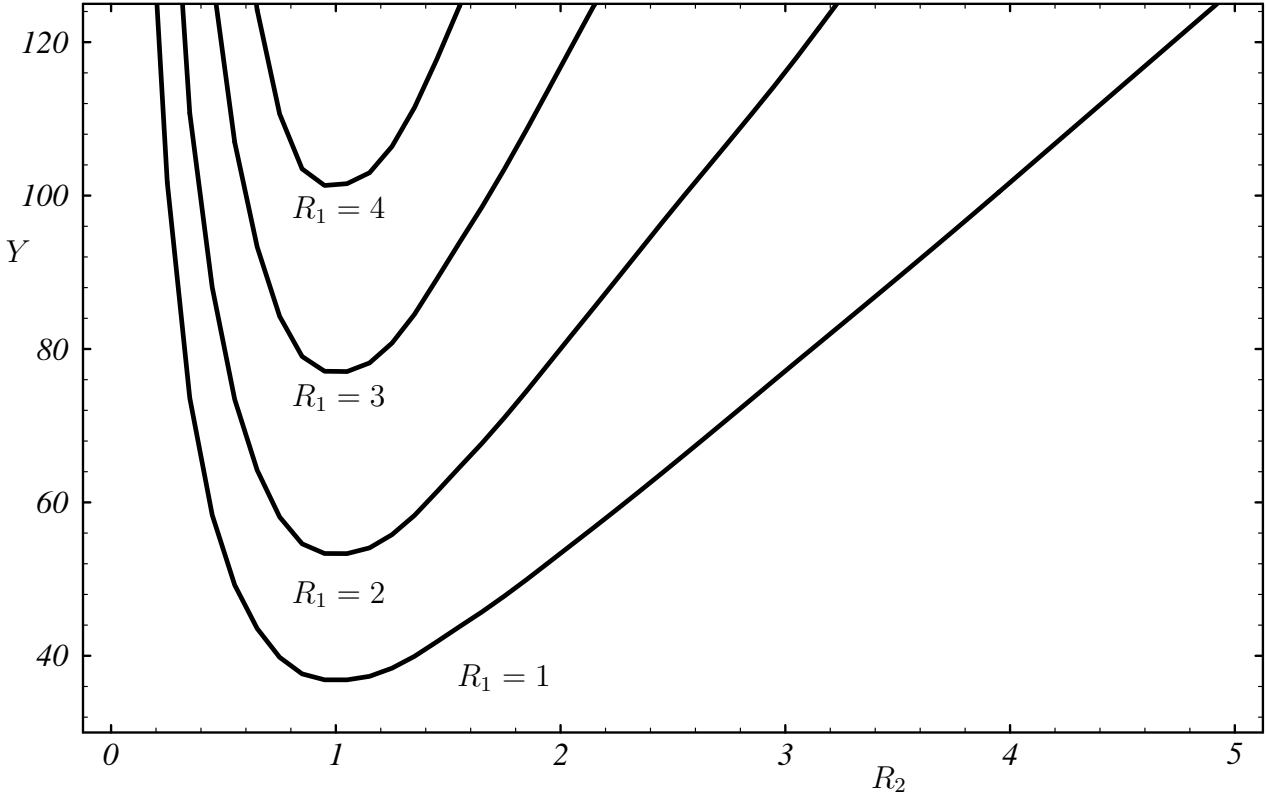}
\caption{Plots of the universal thresholds $Y(R_1,R_2)$ as a function
of $R_2$ for $R_1=1,2,3,4$.}
\vskip -1.5cm\hskip 11cm $R_2$
\vskip -7.5cm\hskip -.5cm $Y$
\vskip -1.1cm\hskip 3.3cm $R_1=4$
\vskip 2.cm\hskip3.3cm$R_1=3$
\vskip 2.1cm\hskip3.3cm$R_1=2$
\vskip .6cm\hskip 5.5cm$R_1=1$
\vskip 2.5cm
\end{figure}
\end{centering}

Let us now analyze the effect of the universal thresholds on the
unification
scale of the low-energy effective field theory. In the model under
consideration, the $E_8\times E_6$ gauge group couplings will run
according to (\ref{one}) with $c_\alpha$ and
$Y(R_1,R_2)$ given by (\ref{con}) and (\ref{yyr}) respectively. There
is some arbitrariness in the definition of the unification scale
$M_U$
but in the specific model
there is a manifest way to define it \cite{ILR}, in the ${\overline
{DR}}$
scheme:
\begin{equation}
M_U(R_1,R_2)  =  \frac{M_p \, g_U}
{{2\pi} \left|\eta\left(i\frac{R_2}{R_1}\right)\right|^2
\left|\eta\left(i R_1 R_2\right)\right|^2 R_2}\times\frac{1}{
\sqrt{1 + g_U^2\frac{Y(R_1,R_2)}{16\pi^2}}}\ ;
\label{dmu}
\end{equation}
here
$g_U\equiv g_\alpha(M_U)=
g_s/\sqrt{1-g_s^2\frac{Y}{16\pi^2}}\,$ for any group factor, and we
have
used
explicitly (\ref{three}) in order to express the unification scale in
terms of the effective field theory parameters $M_p$ and $g_U$.
The last factor in (\ref{dmu}) is due to the existence
of the universal terms which lead to a shift of the dilaton field in
order to reabsorb the universal contributions into the string
coupling.
It is interesting to observe that since $Y(R_1,R_2) > 0$ this extra
factor
always gives  a lower unification scale with respect
to the case where these terms are neglected.
One can consider the
minimum
value of the unification scale $M_U^{\scriptstyle\rm min}$
with respect to the radii $R_1$, $R_2\,$. Since $M_U(R_1,R_2)$
possesses target-space duality properties,
it has an extremum at $R_1 = R_2 = 1$ independently of the value of
$g_U$.
On the other hand the first
factor in (\ref{dmu}) monotonically increases for  radii moving away
from
the self-dual point, while the second one monotonically decreases.
In the asymptotic limit $R_1 = R_2 = R \gg 1$,
the universal thresholds become large and the $g_U$ dependence
of $M_U$ cancels between the two factors of (\ref {dmu}). Moreover,
the first factor
dominates and
using (\ref{yas}) we find that $M_U$ grows exponentially:
$M_U\sim 0.78\times
M_P \times  \frac{e^{\frac{\pi}{6} R^2}}{R^2}$. Numerical evaluation
of
(\ref{dmu})  shows that   for
perturbative values of $g_U$ the $R_1 = R_2 = 1$ extremum
is a minimum and that there is no other minimum apart from that.
Thus for the specific   model, in the ${\overline {DR}}$
scheme,
\begin{equation}
M_U^{\scriptstyle\rm min} = \frac{M_p\, g_U}
{{2\pi}\left|\eta\left(i\right)\right|^4}\times\frac{1}
{\sqrt{1 + g_U^2\frac{Y(1,1)}{16\pi^2}}}\ .
\label{mumin}
\end{equation}
Using explicit values for the various quantities that appear in this
formula we obtain
\begin{equation}
M_U^{\scriptstyle\rm min}
\approx 5.56\times 10^{17}\times g_U\times
\frac{1}
{\sqrt{1 + 0.23 \times g_U^2}}
\ {\rm GeV}
\end{equation}
where the last factor in the product represents the effect of the
universal thresholds.

We would like to conclude this note with a few comments.
We have determined the complete one-loop threshold corrections
(see (\ref{eleven})) for general symmetric-orbifold models, in the
$\overline{DR}$ scheme, by using a method introduced in
\cite{KK} that allows us to handle the infra-red problems.
The group-factor dependent parts of these thresholds
were obtained previously following a different procedure
\cite{VK, DKL}. The two results for the group-factor
dependent contributions (constant and moduli-dependent) are
in agreement, when evaluated within the same ultra-violet
renormalization scheme.
Put differently, this shows that the relation between the
running gauge couplings of the low-energy field theory in the
$\overline{DR}$ scheme and the string coupling does not depend
on the infra-red regularization prescription. This amounts to
the decoupling of the (infinite tower of) massive states and
allows for an unambiguous definition of string effective theory.
Such a conclusion could not have been drown without using
a consistent infra-red regulator.
Although our result has been established in the framework of an
infra-red regulator induced by a particular $N=4$ four-dimensional
curved background, we would have reached the same conclusions
within any other background possessing similar properties,
such as those listed in \cite{BAC}.
\newpage
\begin{centering}
\begin{figure}
\epsfxsize=14.5cm
\epsfbox[40 170 560 620]{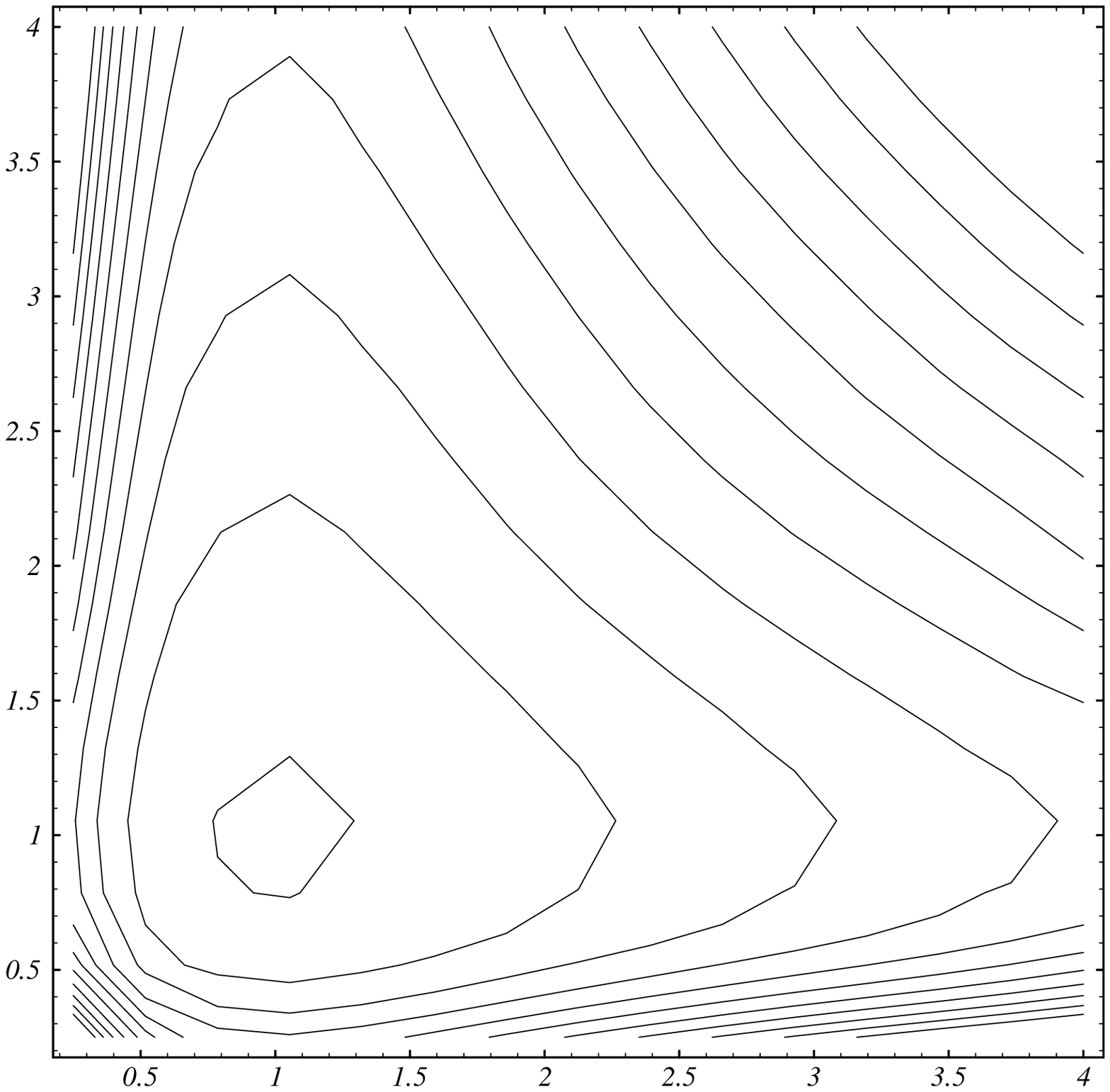}
\caption{ Contour plots of the universal thresholds $Y(R_1,R_2)$
as a function of the internal radii $R_1$ and $R_2$.}
\vskip -1.5cm\hskip11.cm$R_1$
\vskip-11.cm\hskip 0.5cm$R_2$
\vskip 6.5cm\hskip5.3cm $ $
\vskip -.6cm\hskip5.cm $40$
\vskip -2.2cm\hskip6.5cm $60$
\vskip -1.8cm\hskip7.3cm $80$
\vskip -2.2cm\hskip8.7cm $120$
\vskip -1.8cm\hskip10cm $160$
\vskip -1.8cm\hskip10.95cm $200$
\vskip 12.cm
\end{figure}
\end{centering}

Going beyond what has been achieved in previous studies,
we have determined and calculated the moduli-dependent
universal part $Y(T,U)$ of the thresholds for the symmetric
$Z_2\times Z_2$ orbifold model, eq. (\ref{twelve}).
We have found that  $Y(R_1,R_2)$
is strictly positive with a
minimum  $Y^{\scriptstyle\rm min}  \approx 36.4$ at the self-dual
point $R_1=R_2=1$, and is monotonically growing away from it.
For large radii the asymptotic form of the universal
term is $Y(R,R)\sim 18.85\times R^2+17.49 \times R^{-2}$.
It would be interesting to compute the universal thresholds in models
such as the $Z_3$ orbifold, where the $N=1$ sectors do contribute,
and check whether this contribution is indeed as small as one
generally
believes by looking at the group-factor dependent terms \cite{NIL}.

Finally, we have studied the effect of
the universal thresholds on the unification scale of the underlying
field theory.
As we stressed in the introduction, the universal thresholds cannot
be
reabsorbed into a redefinition of the coupling constant without
affecting
the relation between the Planck scale and the unification scale,
and we have actually found that the existence of these terms leads
to a
decrease of that scale. The minimum of the unification
scale is obtained for radii at the self-dual point but the specific
value
depends on the value of the effective field theory gauge coupling
$g_U$
at this scale. For example for $g_U^2=\frac{1}{2}$ we have a
$5\%$ decrease while for $g_U^2=1$ we can reach $10\%$, with
respect
to the case where these corrections are not taken into account.
Of course
one could argue that this unification scale might be lowered further
in some model other than the $Z_2\times Z_2$ orbifold.
However, one should be aware
that the above scale concerns the
$E_8 \times E_6$
symmetry breaking and that one has somehow
to introduce a lower scale where
$E_6$ breaks down to some subgroup, eventually leading to the
standard model.
In order to describe such a realistic situation in the framework of
strings,
it seems difficult to avoid the introduction of $y$-fields and Wilson
lines
\cite{KPR, NIL}.
Those will enhance the moduli space and allow for a better
exploration of the various symmetry-breaking possibilities.

\centerline{\bf Acknowledgements}
We would like to thank E. Kiritsis and
%especially
C. Kounnas for
stimulating   discussions. One of us (J.R.) would like to thank the
CERN
Theory Division  for hospitality and  acknowledges financial
support from the EEC contract ref. ERBCHBGCT940634.

\end{document}